\documentstyle[prd,aps,eqsecnum,epsf,array]{revtex}

\newcommand{\be}{\begin{equation}}
\newcommand{\ee}{\end{equation}}

\newcommand{\bqn}{\begin{eqnarray}}
\newcommand{\eqn}{\end{eqnarray}}

\begin{document}

\title{A minimal size for granular superconductors}
\author{L.M. Abreu , A. P. C. Malbouisson , I. Roditi }
\address{Centro Brasileiro de Pesquisas F\'\i sicas - CBPF/MCT,\\ Rua
Dr.Xavier Sigaud 150\\
22290-180, Rio de Janeiro, RJ, Brazil}
\date{\today}
\maketitle

\begin{abstract}
{\bf Abstract} We investigate the minimal size of small superconducting grains by means of a Ginzburg-Landau model confined to a sphere of radius $R$. This
model is supposed to describe a material in the form of a ball, whose
transition temperature when presented in bulk form, $T_{0}$, is known.
We obtain an equation for the critical temperature as a function of $R$ and
of $T_{0}$, allowing us to arrive at the
minimal radius of the sphere below which no superconducting transition 
exists.\newline
\vspace{0.34cm}\noindent PACS Number(s): 11.10.-z, 74.20.-z, 74.81.Bd
\end{abstract}

\setlength{\topmargin}{-2ex}

\section{Introduction}

Recently, experimental results on small metallic grains \cite{grains} led to an important effort on theoretical investigations of superconducting pairing correlations in nanograins \cite{delft}. Many of the present developments in this area are based on the exact solution of the discrete BCS model \cite{RBCS}. This model is described by a Hamiltonian for a discrete set of doubly-degenerate energy levels containing a pairing interaction for the scattering of pairs of electrons at levels next to each other. One of the fundamental questions addressed by these works has been stated long ago \cite{Anderson} and was restated in the following form \cite{grains}:
{\it What is the lower size limit for the existence of superconductivity in small grains?}

In the present letter we present an answer for the above question from another point of view, stemming from the field theory framework of the Ginzburg-Landau model. Our approach consists in considering an Euclidean massive $(\lambda \varphi ^{4})_{D}$ model describing a system constrained to be confined to a sphere of radius R. The rationale for our procedure is twofold, on one hand the Ginzburg-Landau model provides a well established and elegant theory of the phenomenology of the superconducting state. On the other hand, for Euclidean field theories, temperature, understood as  imaginary time, and spatial coordinates share the same footing, which allows us to apply to the spatial 
coordinates a compactification mechanism using the Matsubara formalism
\cite{Ademir,JMario,Gino1}. We emphasize that we are considering an Euclidean field theory in $D$ {\it  
purely spatial} dimensions, so we are {\it not} working in the framework of
finite temperature field theory. Here, temperature is introduced in the
mass term of the Hamiltonian 
(\ref{Lagrangeana}) below by means of the usual Ginzburg-Landau prescription.

We must stress here that we do not pretend to attain the degree of understanding coming from microscopic theories like BCS \cite{BCS} or the reduced BCS \cite{RBCS}, nevertheless we believe that our perspective can shed some light on this interesting question.    

We consider the
Ginzburg-Landau Hamiltonian density, 
\begin{equation}
{\cal H}=\frac{1}{2}\left( {\bf \nabla }\varphi \right) ^{2}+\frac{1}{2}%
m_{0}^{2}\varphi {^{2}}+\frac{u}{4!}\varphi ^{4},  \label{Lagrangeana}
\end{equation}
in Euclidean $3$-dimensional space, where $u$ is the coupling constant, and $
m^{2}_{0}=\alpha (T-T_{0})$ is the squared bare mass ($T_{0}$ being the bulk transition
temperature of the superconductor and $\alpha >0$). 
From a physical point of view, introducing temperature by means of the mass term in this
Hamiltonian, this should correspond to a spherical sample of material. We
investigate the behaviour of the system as a function of the radius $R$ of the confining 
sphere, in the approximation in which gauge fluctuations are neglected, and using spherical coordinates. Under these conditions one can write the generating functional of the correlation functions as, 
\begin{equation}
{\cal Z}=\int {\cal D}\varphi ^{\ast }{\cal D}\varphi exp\left(
-\int_{0}^{R}dr\int d\Omega \;{\cal H}(\varphi ,\nabla \varphi )\right) ,
\label{Z}
\end{equation}
with the field $\varphi (r,\theta ,\phi )$ satisfying the condition of
confinement along the $r$-axis, $\varphi (r=0)\;=\;\varphi (r=R)\;=\;0$.
These conditions of confinement of the $r$-dependence of the field to a
sphere of radius $R$, permit us to proceed with respect to the $r$%
-coordinate, in a manner analogous as it is done in the imaginary-time
Matsubara formalism in field theory. The Feynman rules should be modified
following the prescription, 
\begin{equation}
\int \frac{dk_{r}}{2\pi }\rightarrow \frac{1}{R}\sum_{n=-\infty }^{+\infty
}\;,\;\;\;\;\;\;k_{r}\rightarrow \frac{2n\pi }{R}\equiv \omega _{n}.
\label{Matsubara}
\end{equation}

The letter is organized as follows. In the next section we calculate the 
compactified effective potential obtaining its R-dependence. The suppression
of superconductivity as a function of the grain radius is shown in section 3.
Section 4 present some concluding remarks. 

\section{The $R$-dependent effective potential}

We start from the expression for the one-loop contribution
to the effective potential in absence of boundaries in spherical
coordinates $U_{1}(\phi_{0} ,R=\infty )$, where in order to deal with dimensionless quantities in the required regularization procedures, we introduce parameters $c^{2}=m^{2}/4\pi ^{2}\mu
^{2}$, $(R\mu )^{2}=a^{-1}$, $g=(u/8\pi ^{2})$, $(\varphi _{0}/\mu )=\phi
_{0}$, $\varphi _{0}$ being the normalized vacuum expectation value of
the field (the classical field) and $\mu $ a mass scale. In terms of
these parameters,
 
\begin{equation}
U_{1}(\phi_{0} ,R=\infty )=\mu ^{3}\sum_{s=1}^{\infty }\frac{(-1)^{s+1}}{2s}%
g^{s}\phi _{0}^{2s}\int \frac{d\;\Omega \;k_{r}^{2}d\;k_{r}}{%
(k_{r}^{2}+c^{2})^{s}}\,,  \label{potefet0}
\end{equation}
here $m$ is the {\it renormalized} mass and the integration over the solid
angle $d\;\Omega $ gives simply a factor $4\pi $. Performing the
replacements defined in (\ref{Matsubara}), we obtain the boundary-dependent ($R$-dependent)
one-loop contribution to the effective potential in the form, 
\begin{eqnarray}
U_{1}(\phi_{0} ,R) &=&4\pi \mu ^{3}a^{3/2}\sum_{s=1}^{\infty }\frac{(-1)^{s+1}}{%
2s}g^{s}\phi _{0}^{2s}\times   \nonumber \\
&&\times \sum_{n=-\infty }^{+\infty }\frac{n^{2}}{(an^{2}+c^{2})^{s}}.
\label{potefet1}
\end{eqnarray}
This equation can be rewritten using the property 
\begin{equation}
\sum_{n=-\infty }^{+\infty }\frac{n^{2}}{(an^{2}+c^{2})^{s}}=-\frac{1}{s-1}%
\frac{d}{da}A_{1}^{c^{2}}(s-1;a),  \label{prop0}
\end{equation}
where $A_{1}^{c^{2}}(s-1;a)$ is one of the Epstein-Hurwitz $zeta$ -functions 
\cite{Elizalde} defined by, 
\begin{equation}
A_{d}^{c^{2}}(\nu ;a_{1},...,a_{d})=\sum_{\{n_{i}\}=-\infty }^{+\infty }%
\frac{1}{(a_{1}n_{1}^{2}+...+a_{d}n_{d}^{2}+c^{2})^{\nu }},  \label{zeta0}
\end{equation}
valid for $Re(\nu )>\;d/2$ (in our case $Re(s)>\;1/2$). Then Eq.(\ref
{potefet1}) becomes, 
\begin{eqnarray}
U_{1}(\phi_{0} ,R) &=&-4\pi \mu ^{3}a^{3/2}\sum_{s=1}^{\infty }\frac{(-1)^{s+1}}{%
2s(s-1)}g^{s}\phi _{0}^{2s}\times   \nonumber \\
&&\times \frac{d}{da}A_{1}^{c^{2}}(s-1;a).  \label{potefet2}
\end{eqnarray}
The Epstein-Hurwitz $zeta$-function can be extended to the whole complex $s$
-plane generalizing as is done in \cite{Ademir} the mode sum regularization
procedure described in Ref.\cite{Elizalde}. We write, 
\begin{eqnarray}
A_{d}^{c^{2}}(\nu ;a_{1},...,a_{d}) &=&\frac{1}{c^{2\nu }}%
+2\sum_{i=1}^{d}\sum_{n_{i}=1}^{\infty }(a_{i}n_{i}^{2}+c^{2})^{-\nu }+ 
\nonumber \\
&&2^{2}\sum_{i<j=1}^{d}\sum_{n_{i},n_{j}=1}^{\infty
}(a_{i}n_{i}^{2}+a_{j}n_{j}^{2}+c^{2})^{-\nu }+\cdots   \nonumber \\
&&+2^{d}\sum_{n_{1},...,n_{d}=1}^{\infty }(a_{1}n_{1}^{2}+\cdots
+a_{d}n_{d}^{2}+c^{2})^{-\nu }.  \label{zeta1}
\end{eqnarray}
Using the identity, 
\begin{equation}
\frac{1}{\Delta ^{\nu }}=\frac{1}{\Gamma (\nu )}\int_{0}^{\infty }dt\;t^{\nu
-1}e^{-\Delta t},
\end{equation}
we get, 
\begin{eqnarray}
A_{d}^{c^{2}}(\nu ;a_{1},...,a_{d}) &=&\frac{1}{\Gamma (\nu )}%
\int_{0}^{\infty }dt\;t^{\nu -1}e^{-c^{2}t}\left[ 1+2%
\sum_{i=1}^{d}T_{1}(t,a_{i})+\right.   \nonumber  \label{zeta1} \\
&&\left. +2^{2}\sum_{i,j=1}^{d}T_{2}(t,a_{i},a_{j})+\cdots
+2^{d}T_{d}(t,a_{1},...,a_{d})\right] ,
\end{eqnarray}
where, 
\begin{eqnarray}
T_{1}(t,a_{i}) &=&\sum_{n_{i}=1}^{\infty }e^{-a_{i}n_{i}^{2}t}\;,  \label{T1}
\\
T_{j}(t,a_{1},...,a_{j})
&=&T_{j-1}(t,a_{1},...,a_{j-1})T_{1}(t,a_{j})\;\;\;\;,j=2,...,d.
\end{eqnarray}
Considering the property of functions $T_{1}$, 
\begin{equation}
T_{1}(t,a_{i})=-\frac{1}{2}+\sqrt{\frac{\pi }{a_{i}t}}\left[ \frac{1}{2}+S(%
\frac{\pi ^{2}}{a_{i}t})\right] ,  \label{T2}
\end{equation}
where 
\begin{equation}
S(x)=\sum_{n=1}^{\infty }e^{-n^{2}x},  \label{S}
\end{equation}
we can notice that the surviving terms in Eq.(\ref{zeta1}) are proportional
to $(a_{1}\cdots a_{d})^{-(1/2)}$. Therefore we find, 
\begin{eqnarray}
A_{d}^{c^{2}}(\nu ;a_{1},...,a_{d}) &=&\frac{\pi ^{\frac{d}{2}}}{\sqrt{%
a_{1}\cdots a_{d}}}\frac{1}{\Gamma (\nu )}\int_{0}^{\infty }dt\;t^{(\nu -%
\frac{d}{2})-1}e^{-c^{2}t}  \nonumber  \label{zeta3} \\
&&\times \left[ 1+2\sum_{i=1}^{d}S(\frac{\pi ^{2}}{a_{i}t}%
)+2^{2}\sum_{i<j=1}^{d}S(\frac{\pi ^{2}}{a_{i}t})S(\frac{\pi ^{2}}{a_{j}t}%
) \right. \nonumber \\
& & \left. +\cdots +2^{d}\prod_{i=1}^{d}S(\frac{\pi ^{2}}{a_{i}t})\right] .\nonumber \\
\label{zeta3}
\end{eqnarray}
Inserting in Eq.(\ref{zeta3}) the explicit form of the function $S(x)$ in
Eq.(\ref{S}) and using the following representation for Bessel functions of
the third kind, $K_{\nu }$, 
\begin{equation}
2(a/b)^{\frac{\nu }{2}}K_{\nu }(2\sqrt{ab})=\int_{0}^{\infty }dx\;x^{\nu
-1}e^{-(a/x)-bx},  \label{K}
\end{equation}
we obtain after some long but straightforward manipulations, 
\begin{eqnarray}
A_{d}^{c^{2}}(\nu ;a_{1},...,a_{d}) &=&\frac{2^{\nu -\frac{d}{2}+1}\pi
^{2\nu -\frac{d}{2}}}{\sqrt{a_{1}\cdots a_{d}}\,\Gamma (\nu )}\left[ 2^{\nu -%
\frac{d}{2}-1}\Gamma (\nu -\frac{d}{2})(\frac{m}{\mu })^{d-2\nu }\right.  
\nonumber  \label{zeta4} \\
&&\left. +2\sum_{i=1}^{d}\sum_{n_{i}=1}^{\infty }(\frac{m}{\mu ^{2}L_{i}n_{i}%
})^{\frac{d}{2}-\nu }K_{-\nu +\frac{d}{2}}(mL_{i}n_{i})+\cdots \right.  
\nonumber \\
&&\left. +2^{d}\sum_{n_{1},...,n_{d}=1}^{\infty }(\frac{m}{\mu ^{2}\sqrt{%
L_{1}^{2}n_{1}^{2}+\cdots +L_{d}^{2}n_{d}^{2}}})^{\frac{d}{2}-\nu } \times \right. \nonumber \\
&& \left. \times K_{-\nu +%
\frac{d}{2}}(m\sqrt{L_{1}^{2}n_{1}^{2}+\cdots +L_{d}^{2}n_{d}^{2}})\right] .
\nonumber \\
&&
\end{eqnarray}
For $d=1$, taking $\nu =s-1$, and identifying the compactified dimension
with the radial coordinate, we obtain in our case from Eq.(\ref{potefet2})
the one-loop correction to the effective potential, 
\begin{eqnarray}
U_{1}(\phi_{0} ,R) &=&-4\pi \mu ^{3}a^{3/2}\sum_{s=1}^{\infty }\frac{(-1)^{s+1}}{%
2s(s-1)}g^{s}\phi _{0}^{2s}\frac{2^{s-1/2}\pi ^{2s-5/2}}{\Gamma (s-1)}\times 
\nonumber \\
&&\times \frac{d}{da}\left( \frac{1}{\sqrt{a}}\left[ 2^{s-5/2}\Gamma (s-%
\frac{3}{2})(\frac{m}{\mu })^{3-2s}+\right. \right.   \nonumber \\
&&\left. \left. +2\sum_{n=1}^{\infty }(\frac{m}{\mu ^{2}nR})^{\frac{3}{2}%
-s}K_{s-\frac{3}{2}}(mnR)\right] \right),  \label{potefet4}
\end{eqnarray}
or, remembering $a^{-1}=(R\mu )^{2}$ and $\phi_{0}= \varphi_{0}/ \mu$, 
\begin{eqnarray}
U_{1}(\varphi_{0} ,R) &=&\sum_{s=1}^{\infty }\frac{(-1)^{s+1}}{s\Gamma (s)}%
2^{s+1/2}\pi ^{2s-3/2}g^{s}\varphi _{0}^{2s}\times   \nonumber \\
&&\times \left[ 2^{s-7/2}\Gamma (s-\frac{3}{2})\; 
m ^{3-2s}+\sum_{n=1}^{\infty }\left( \frac{m}{nR}\right) ^{\frac{3}{2}%
-s}K_{-s+\frac{3}{2}}(mnR)\right.   \nonumber \\
&&\left. +R\frac{d}{dR}\sum_{n=1}^{\infty }\left( \frac{m}{nR}
\right) ^{\frac{3}{2}-s}K_{-s+\frac{3}{2}}(mnR)\right] .  \label{potefet5}
\end{eqnarray}
where $K_{-s+\frac{3}{2}}$ are Bessel functions of the third kind.

\section{Critical behaviour}

As we are analizing the case where the field has only one component we can neglect the $R$-dependence of the coupling constant, that is we consider $u$ as the {\it renormalized} coupling constant. In this case, it is enough for us to use only one renormalization
condition, 
\begin{equation}
\frac{\partial ^{2}}{\partial \varphi_{0} ^{2}}U_{1}(\varphi_{0} ,R)|_{\varphi
_{0}=0}=m^{2}.  \label{renorm1}
\end{equation}
Notice that we are using a modified minimal subtraction scheme \cite{MMS}, where the
mass (and coupling constant, if it is the case) counter-terms are for {\it %
even} space dimension $D$, poles of $Gamma$-functions at the physical values
of $s$ ($s=1$ for the mass, $s=2$ for the coupling constant). For arbitrary $%
D$, we would have in the $R$-independent term in Eq.(\ref{potefet5}),
instead of the factor $\Gamma (s-\frac{3}{2})$, a factor $\Gamma (s-\frac{D}{%
2})$, which would generate a pole at $s=1$ for even dimensions. This polar
term should be subtracted, giving the $R$-dependent correction to the
renormalized mass proportional to the regular part of the analytical
extension of the Epstein-Hurwitz $zeta$-function in the neighbourhood of the
pole at $s=1$. For {\it odd} space dimensions, as for our case $D=3$, there
are no poles of $Gamma$-functions, but in order to have a coherent procedure
in any dimension, we also subtract the corresponding term, performing a
finite renormalization. Thus the $R$-dependent renormalized mass, at
one-loop approximation, is given by 
\begin{eqnarray}
m^{2}(R) &=&m_{0}^{2}+\frac{u }{3\sqrt{2}\pi ^{3/2}}\left[
\sum_{n=1}^{\infty }\left( \frac{m_{0}}{nR}\right) ^{1/2}K_{\frac{1}{2}%
}(nRm_{0})+\right.   \nonumber \\
&&\left. +R\frac{d}{dR}\sum_{n=1}^{\infty }\left[ \frac{m_{0}}{nR}\right]
^{1/2}K_{\frac{1}{2}}(nRm_{0})\right]   \label{massaR}
\end{eqnarray}

On the other hand, if we start in the ordered phase, the model exhibits
spontaneous symmetry breaking, but for sufficiently small values of $T^{-1}$
and $R$ the symmetry is restored. We can define the critical curve $%
C(T_{c},R)$ as the curve in the $T\times R$ plane for which the inverse
squared correlation length , $\xi ^{-2}(T,R,\varphi _{0})$, vanishes in the $%
R$-dependent gap equation, 
\begin{eqnarray}
&&\left. \xi ^{-2}=m_{0}^{2}+u {\bf \varphi }_{0}^{2}+\right.   \nonumber
\\
&&\;\;\;+\frac{u }{6R}\sum_{n=-\infty }^{\infty }\int \frac{d\Omega }{%
(2\pi )^{2}}\;\frac{\omega _{n}^{2}}{\omega _{n}^{2}+\xi ^{-2}}\,,
\label{gap}
\end{eqnarray}
where ${\bf \varphi }_{0}$ is the normalized vacuum expectation value of the
field (different from zero in the ordered phase). In the disordered phase,
in particular in the neighborhood of the critical curve, $\varphi _{0}$
vanishes and the gap equation reduces to a $R$-dependent Dyson-Schwinger
equation, 
\begin{eqnarray}
m^{2}(T,R) &=&m_{0}^{2}(T)+\frac{u }{6R}  \nonumber \\
&&\times \sum_{n=-\infty }^{\infty }\int \frac{d\Omega }{(2\pi )^{2}}\;\frac{%
\omega _{n}^{2}}{\omega _{n}^{2}+m^{2}(T,R)}.  \label{gap1}
\end{eqnarray}

After steps analogous to those leading from Eq.(\ref{potefet1}) to Eq.(\ref
{massaR}), Eq.(\ref{gap1}) can be written in the form 
\begin{eqnarray}
m^{2}(T,R) &=&m_{0}^{2}+\frac{u}{3\sqrt{2}\pi ^{3/2}}\times   \nonumber
\\
&&\times \left[ \sum_{n=1}^{\infty }\left( \frac{m(T,R)}{nR}\right) ^{1/2}K_{%
\frac{1}{2}}(nRm(T,R))+\right.   \nonumber \\
&&\left. +R\frac{d}{dR}\sum_{n=1}^{\infty }\left( \frac{m(T,R)}{nR}\right)
^{1/2}K_{\frac{1}{2}}(nRm(T,R))\right] .  \nonumber \\
&&  \label{massaR1}
\end{eqnarray}
If we limit ourselves to the neighborhood of criticality, $m^{2}(T,L)\approx
0$, we may investigate the behavior of the system by using in Eq.(\ref
{massaR1}) an asymptotic formula for small values of the argument of Bessel
functions, 
\begin{equation}
K_{\nu }(z)\approx \frac{1}{2}\Gamma (\nu )\left( \frac{z}{2}\right) ^{-\nu
}\;\;\;(z\sim 0).  \label{K}
\end{equation}
Performing the derivative in Eq.(\ref{massaR1}) with the help of the
formula, 
\begin{equation}
\frac{d}{dz}K_{v}(z)=vz^{-1}K_{v}(z)-K_{v+1}(z),  \label{der}
\end{equation}
and using the tabulated values $\Gamma (1/2)=\sqrt{\pi }$, $\Gamma (3/2)=%
\sqrt{\pi }/2$, we obtain after some manipulations in the neighbourhood of
criticality, 
\begin{equation}
m^{2}(T,R)\approx m_{0}^{2}+\frac{u \zeta (1)}{6\pi R},  \label{mDysoncr}
\end{equation}
where $\zeta (z)$ is the Riemann $zeta$-function defined for $z>1$.

Eq.(\ref{mDysoncr}) is of course meaningless as it stands. In order to obtain a critical curve and give a meaning to eq.(\ref{mDysoncr}), we perform an analytic continuation of the $zeta$-function $%
\zeta (z)$ to values of the argument $z\leq 1$, by means of its reflection
property, 
\begin{equation}
\zeta (z)=\frac{1}{\Gamma (z/2)}\Gamma (\frac{1-z}{2})\pi ^{z-\frac{1}{2}%
}\zeta (1-z),  \label{extensao}
\end{equation}
which defines a meromorphic function having only one simple pole at $z=1$.
The physical interpretation is achieved through a mass renormalization
procedure that can be done as follows: remembering the formula 
\begin{equation}
\lim_{z\rightarrow 1}\left[ \zeta (z)-\frac{1}{1-z}\right] =\gamma ,
\label{extensao1}
\end{equation}
where $\gamma \cong 0.57216$ is the Euler constant, the {\it %
renormalized} mass $\bar{m}$ is then defined as 
\begin{eqnarray}
\bar{m}^{2}(T,R) &=&\lim_{z\rightarrow 1_{+}}\left[ m^{2}(T,R)-\frac{u }{%
6\pi R(z-1)}\right]   \nonumber \\
&=&\alpha (T-T_{0})+\frac{u \gamma }{6\pi R}.  \label{massaRR}
\end{eqnarray}
Taking this {\it renormalized} mass equal to zero leads to the critical
temperature, given as a function of the radius $R$ by, 
\begin{equation}
T_{c}=T_{0}-\frac{u \gamma }{6\pi R}.  \label{critica}
\end{equation}
In Eq.(\ref{critica}), $T_{0}$ corresponds to the transition temperature for
the material in absence of boundaries ($R\rightarrow \infty $), that is,
to the bulk transition temperature. We see then that, in a spherical sample
made of the same material, the critical temperature is diminished by a
quantity proportional to the inverse of its radius. Also, we see that there
is a minimal radius $R^{(0)}$ below which superconductivity is
suppressed. Identifying the usual $3$-dimensional Ginzburg-Landau parameter $
\beta =u$, the minimal radius is given by, 
\begin{equation}
R^{(0)}=\frac{\gamma \beta }{6\pi \alpha T_{0}}.  \label{supressao}
\end{equation}

\section{Concluding remarks}

In this letter, within the framework of the Ginzburg-Landau model, we were able to describe a spherical system, a grain, and we showed that, below a specific size, superconductivity is suppressed. Our method made use of effective potential and dimensional renormalization techniques supplemented by the Matsubara prescription (\ref{Matsubara}) which give us a structure to work in a spherical geometry. We did not attempt to deal with the microscopic properties of ultrasmall metallic grains, however we believe that our approach can pave the way to further studies of the superconducting state in different geometries, notably through Monte Carlo and variational techniques.

\section{Acknowledgements}
This work has been supported by the Brazilian agency CNPq (Brazilian 
National Research Council). One of us, IR, also thank Pronex/MCT for 
partial support.

\bigskip

\end{document}